\documentclass[aps,amsmath,amssymb,preprint]{revtex4} 

\usepackage{graphicx}
\usepackage{dcolumn}
\usepackage{bm}
\usepackage{natbib}

\begin{document}

\bibliographystyle{naturemag} 

\title{Inverse spin Hall effect in a complex ferromagnetic oxide heterostructure}

\author{M. Wahler$^{1}$}%
\author{N. Homonnay$^{1}$}%
\author{T. Richter$^{1}$}%
\author{A. M\"{u}ller$^{1}$}%
\author{B. Fuhrmann$^{2}$}%
\author{G. Schmidt$^{1,2,*}$}%

\affiliation{%
$^1$Institut f\"{u}r Physik, Martin-Luther-Universit\"at Halle-Wittenberg, 06120 Halle, Germany\\
$^2$Interdisziplin\"{a}res Zentrum f\"{u}r Materialwissenschaften, Martin-Luther-Universit\"at Halle-Wittenberg, 06120 Halle, Germany\\
$^*$Correspondence to G. Schmidt, email: georg.schmidt@physik.uni-halle.de}

\begin{abstract}
Complex oxide heterostructures are hot candidates for post CMOS multi-functional devices. Especially in spintronics applications ferromagnetic oxides may play a key role because they can exhibit extraordinary high spin polarization. Indeed, there are already plenty of examples in spintronics, notably in the area of spin pumping and inverse spin Hall effect (ISHE) \cite{Azevedo2011, Czeschka2011, Hahn2013, Obstbaum2014}. Although complex oxides have been used in these experiments as a source of spin currents, they have never been demonstrated to act as a spin sink that exhibits ISHE. Here we show that in a heterostructure consisting of La$_{0.7}$Sr$_{0.3}$MnO$_{3}$ (LSMO) and SrRuO$_{3}$ (SRO) the low temperature ferromagnet SRO can act as a spin sink and exhibit a sizeable ISHE which persists even below its Curie temperature.
This result opens up new possibilities for application of all oxide heterostructures in spintronics and may significantly extend the research on spin Hall effect and related phenomena.
\end{abstract}

\maketitle

The conversion of pure spin currents into charge currents has widely been investigated over the past decade and a large variety of materials have been used in these experiments \cite{Mosendz2010,Ando2013,Hahn2013}. In the prototypical setup a ferromagnet is excited to ferromagnetic resonance using a microwave signal. The resulting precession of the magnetization causes a spin current into an adjacent conducting material \cite{Tserkovnyak2002a}. This spin current cannot be measured directly but first needs to be converted into a charge current or a voltage \cite{Zutic2011,Jungwirth2012}. This so called spin-charge conversion can either be achieved by measuring spin-accumulation using secondary ferromagnetic contacts as spin selective voltage probes \cite{Jedema2002} or using the ISHE \cite{Jungwirth2012,Hahn2013}. Especially the ISHE has been of increasing importance over the last years. Reasons are not only the simplicity of the necessary experimental setup but also the possible application of SHE and ISHE in spintronics. The spin Hall effect can for example be used to increase the efficiency of spin transfer torque for MRAM application \cite{Liu2012} and the ISHE is not only an excellent candidate for measuring pure spin currents but is even under consideration for energy harvesting by converting spin currents generated by thermal gradients into electricity \cite{Bauer2012}. The materials which are used to measure the ISHE are plenty. Most of them are metals like Pt or Au \cite{Mosendz2010} and more recently W \cite{Pai2012} and Ta \cite{Liu2012,Hahn2013} thin films. But also for organic conductors \cite{Ando2013} and inorganic semiconductors \cite{Ando2012,Chen2013a} ISHE has been demonstrated. In view of the multitude of available complex oxides and the prospect of future oxide thin film based electronics \cite{Blamire2009,Mannhart2010,Bibes2011,Ha2011,Opel2012} it is obvious that including a complex oxide into the catalogue of suitable materials would open up new perspectives. Moreover it has been claimed that the ISHE also appears in ferromagnetic materials \cite{Miao2013}. While it is difficult to setup a suitable experiment for ferromagnetic metals it is much easier to realize it in a complex oxide heterostructure. The fact that many ferromagnetic oxides have a Curie temperature below room temperature which is normally considered as a disadvantage can greatly facilitate the experiment, because it allows for switching the magnetization on and off just by changing the temperature, which is not an option for typical metallic ferromagnets where ${T_C}$ is several hundreds of $^\circ$C \cite{Kittel2005a}. For our experiment we find the ideal candidate in the combination of LSMO and SRO, where both materials are ferromagnets at low temperature, however, above $T=155\,{\rm K}$ only the ferromagnetism of the LSMO prevails \cite{Koster2012}.

\begin{figure}[ht]
\centering
\includegraphics[width=0.50\textwidth]{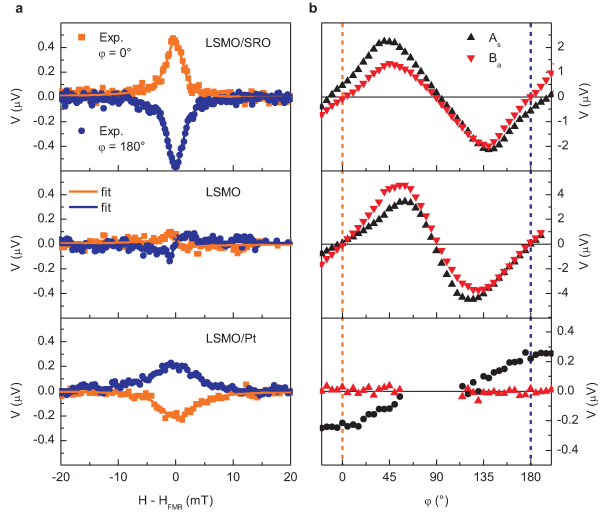}
\caption{\textbf{Separating the ISHE voltage from the AMR-generated voltage. a }, H-dependence of the generated voltage signal for two directions $\varphi=0$ and $\varphi=180^{\circ}$ for the three types of samples. For the LSMO/Pt and the LSMO/SRO samples a symmetric shape of the line is dominant, whereas for the pure LSMO sample, where only AMR contributes to the voltage signal, the symmetric component vanishes. Note that the voltage has opposite sign for SRO and Pt \textbf{b}, Angular dependence of the symmetric and antisymmetric contribution from fits to $V_{DC}=A_s\times L_s(H)+B_a\times L_a(H)+\rm{offset}$. At $\varphi=0$ and $\varphi=180^{\circ}$, both symmetric and antisymmetric contribution become negligible. Nevertheless, for the LSMO/SRO and LSMO/Pt sample a clear symmetric contribution is present which does not contain any contribution from AMR.
\label{pic1}}
\end{figure}

In our experiments we use a total of nine samples all based on a $(30\pm0.4)\,{\rm nm}$ LSMO layer on NdGaO$_{3}$ substrate. Seven of these samples are covered by SRO. The layer thickness are $1.2\,{\rm nm}$, $1.3\,{\rm nm}$, $2.0\,{\rm nm}$, $3.7\,{\rm nm}$, $9.5\,{\rm nm}$, $11.0\,{\rm nm}$ and $22.0\,{\rm nm}$ with error bars of $\pm0.4\,{\rm nm}$ and are named sample 1 to 7, respectively. The two remaining samples are used as reference samples, one without any cap layer on the LSMO and the other covered by $(8\pm1)\,{\rm nm}$ of Pt. These two samples are referred to as reference 1 and reference 2, respectively.
In a first set of measurements on sample 2 and reference 1 and 2 we determine the DC voltage over the sample during microwave excitation at a frequency of $9.6\,\rm{GHz}$ and a temperature of $190\,\rm{K}$ which is above ${\rm T_C}$ of SRO but below the ${\rm T_C}$ of LSMO. The magnetic field $\mu_{0}H_{\rm ext}$ is swept from below to above the resonance field for the LSMO and the measurement is repeated for different in-plane alignments of the magnetic field in steps of $5^\circ$. This set of measurements is performed on all three samples.
For the analysis it is necessary to remove the contributions which stem from RF-rectification due to anisotropic magnetoresistance (AMR) which can occur in a conducting ferromagnet under RF excitation. For certain conditions the precessing magnetization can cause the resistance to oscillate in phase with the induced RF currents, leading to rectification and DC voltages \cite{Mecking2007}. Only when the external field is perpendicular to the RF field no rectification is expected. These AMR-caused voltages need to be carefully separated from the ISHE voltage in order to extract the right result as has been shown by Obstbaum \textit{et al.}\cite{Obstbaum2014}. As a first step we compare the voltage signal generated in the LSMO/SRO heterostructure (sample 2) with that of a single LSMO layer of the same thickness (reference 1). We fit the data by a lorentzian line shape with a symmetric and an antisymmetric part. According to \cite{Obstbaum2014} the ISHE voltage only results in a symmetric contribution, while the signal from AMR can result in both, symmetric and antisymmetric contributions. Fig. \ref{pic1}\textbf{b} shows that for the single LSMO layer symmetric and antisymmetric contributions of finite amplitudes are generated at different magnetization directions in the sample plane. Only when the external magnetic field is directed along the waveguide ($\varphi=0$ or $\varphi=180^\circ$), no DC voltage can be measured as is expected from theory \cite{Mecking2007}. We can thus conclude in reverse that for this alignment of the magnetic field any DC voltage generated in other LSMO based heterostructures does not stem from AMR rectification.

For reference 2 (LSMO/Pt) the Pt has a much lower resistance than the LSMO and represents a short circuit for the AMR generated signal while creating a strong ISHE voltage by itself. The angle dependent measurements (Fig. \ref{pic1}\textbf{b}) confirm that in this sample the maximum DC signal appears for the angle at which no AMR signal could be observed for reference sample 1, as expected from geometric considerations.

For LSMO/SRO heterostructures the result represents a mixture of the two because the SRO has much lower conductivity than Pt and thus a sizeable AMR contribution from the LSMO remains visible (Fig. \ref{pic1}\textbf{b}). For SRO the ISHE has the opposite sign when compared to Pt, which has also previously been observed for Mo \cite{Mosendz2010}, Ta \cite{Liu2012} and W \cite{Pai2012}.

For further analysis we limit ourselves to the amplitude of the symmetric contribution which can be precisely determined at the angle where the AMR signal and thus the antisymmetric contribution vanishes completely as we have shown for the pure LSMO layer (broken line in Fig. \ref{pic1}\textbf{b}). In this case, the spin Hall voltage can be written as
\begin{equation}
	V_{\rm ISHE}\vert_{\varphi=0,H=H_r}=
	\Theta_{\rm SH}
	\frac
		{e}
		{\sigma_{NM}}
	\frac
		{1}
		{2\pi M^{2}_{S}}
	\frac
		{\lambda_{\rm SD}}
		{t_{\rm NM}}
	l
	g^{\uparrow\downarrow}
	\omega 
	h^{2}_{y} 
	\rm{Im}\! 
		\left(
			\chi^{res}_{yy}
		\right)
	\chi^{res}_{zy}
	\;\rm{tanh}\!
		\left(
		 \frac
		 	{t_{\rm NM}}
		 	{2\lambda_{\rm SD}}
		\right)
  {\rm ,}
	\label{VSP_short_Obstbaum2014}
\end{equation}

if out-of-plane excitation is neglected. The formula is derived as described in \cite{Azevedo2011,Obstbaum2014} and contains the conductivity of the SRO ($\sigma_{\rm NM}$), the saturation magnetization of the LSMO layer ($M_{\rm S}$), the length  $l$ over which excitation takes place, the excitation field in y-direction ($h_{y}$), the excitation frequency ($\omega /2\pi$) and the susceptibilities of the precessing magnetization at resonance ($\chi^{res}_{yy}$ and $\chi^{res}_{zy}$).

\begin{figure}[ht]
\centering
\includegraphics[width=0.50\textwidth]{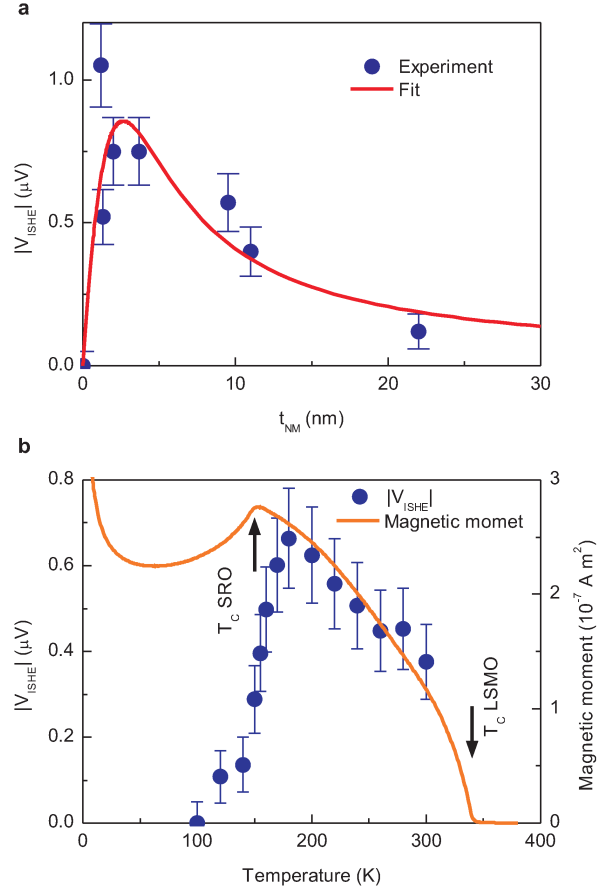}
\caption{\textbf{Quantitative analysis of the voltage generated by ISHE. a,} ISHE voltage as a function of the paramagnetic SRO thickness with a fixed thickness of the ferromagnetic LSMO of $30\,\rm{nm}$ measured at a temperature of  $190\,\rm{K}$. From the fit to equation (\ref{VSP_short}) a spin diffusion length of $2\,\rm{nm}$ is obtained. \textbf{b}, Temperature dependence of $\rm{V}_{\rm{SP}}$. Below $T_{C}({\rm SRO})$ the voltage generated due to ISHE decreases dramatically. Approaching $T_{C}({\rm LSMO})$ the magnetization and therefore the spin Hall voltage decreases.
\label{ThicknessDep}}
\end{figure}

To determine the spin Hall angle $\Theta_{\rm SH}$ of the SRO, the spin-diffusion length $\lambda_{SD}$ of the SRO and the spin-mixing conductance $g^{\uparrow\downarrow}$ of the LSMO/SRO interface must be determined. The latter is obtained by comparing the damping parameter $\alpha$ of the SRO-covered LSMO layer with that of the bare LSMO layer $\alpha_0$ similar to \cite{Azevedo2011}. For this purpose FMR measurements are performed in the frequency range from $2\,\rm{GHz}$ to $25\,\rm{GHz}$ for both samples and the damping parameters are extracted from the line widths. This yields a spin mixing conductance of $g^{\uparrow\downarrow}=(1.1\pm0.3)\times10^{19}\,{\rm m}^{-1}$.
To calculate the spin diffusion length $\lambda_{\rm SD}$ we measure the ISHE voltage on a total of seven samples with varying thickness of the SRO. The results are shown in Fig. \ref{ThicknessDep}\textbf{a}. These values are fit as
\begin{equation}
	V_{\rm ISHE}\propto
			{t_{\rm NM}}^{-1} {\rm tanh}\left(t_{\rm NM}/2\lambda_{\rm SD}\right) {\rm tanh}\left(t_{\rm NM}/\lambda_{\rm SD}\right)
	\label{VSP_short}
\end{equation}

as suggested by \cite{Ando2013} resulting in  $\lambda_{\rm SD}=(1.7\pm0.6)\,{\rm nm}$. In addition we use $\sigma_{\rm NM}=\left(2.0\pm0.5\right)\times10^6\,\Omega^{-1}\rm{m}^{-1}$ from 2-point measurement, $M_{\rm S}=(3.8\pm0.8)\times10^5\,{\rm A}/{\rm m}$ from SQUID-magnetometry, ${\rm l}=600\,\mu{\rm m}$ (width of the signal line), $h_{y}=20\,{\rm A}/{\rm m}$ from DC-current approximation, $\omega /2\pi= 9.6\,{\rm GHz}$ and the susceptibilities calculated from the ferromagnetic resonance and we obtain $\left|\Theta_{\rm SH}\right|=(0.032\pm0.021)$ at $T=190\,{\rm K}$ with the sign being opposite to that of the spin Hall angle of Pt.

In further experiments we measure the temperature dependence of the effect from $100\,\rm{K}$ to $300\,\rm{K}$ to investigate the influence of the ferromagnetism in SRO which has a Curie-temperature of 155 K. Fig. \ref{ThicknessDep}\textbf{b} shows the magnetization of the LSMO/SRO bilayer during cooldown in a field of $2\,{\rm Oe}$ measured by SQUID magnetometry. The change in slope at $T=155\,{\rm K}$ indicates the onset of ferromagnetism in the SRO. SRO is known to have a strong crystalline anisotropy with an out-of-plane easy axis \cite{Koster2012} which may also couple to the LSMO layer. In the same diagram the circles show the corresponding ISHE voltage.
Obviously the ISHE voltage exhibits a maximum at around $T=180\,{\rm K}$. When the temperature is increased starting from $T=180\,{\rm K}$ we observe a decay of the ISHE voltage. This is easily understood from the reduction of the magnetization (and thus the spin polarization) of the LSMO layer which also reduces the spin pumping.

It is quite understandable that the ISHE is reduced upon cooling beyond $T_{C}({\rm SRO})$, however, it is noteworthy that a sizeable effect persists until well below $T_{C}({\rm SRO})$ confirming other experiments that claim ISHE in a ferromagnetic layer \cite{Miao2013}. Note that the vanishing ISHE signal does not even prove that the ISHE itself truly vanishes at lower temperatures. In principle it is possible either that due to its strong out-of-plane anisotropy the SRO may couple to the LSMO and pull the magnetization out of the plane thus changing the geometry of the experiment or that the magnetization vector of the SRO is non-collinear to that of the LSMO and the spin polarization of the current induced by the spin pumping is changed by spin transfer torque. Both effects can make the ISHE undetectable.
Nevertheless, a finite ISHE signal can clearly be observed even beyond the onset of ferromagnetism demonstrating for the first time an ISHE signal in a ferromagnet induced by precessional spin pumping from a second ferromagnet. This unique experiment is greatly facilitated by the possibility of using the temperature as an additional control parameter to effectively switch magnetization on and off, which is hardly possible in typical ferromagnetic metals.

In summary we have shown that SRO can act as a spin sink which exhibits ISHE with a sign opposite to that of Pt. The spin Hall angle is smaller than for Pt and the effect persists even below the Curie temperature of the SRO. The fact that spin pumping and ISHE are present in an all oxide heterostructure opens up new possibilities for future investigation of both effects.

\section*{Methods}
\subsection*{Sample preparation}
Pulsed Laser Deposition (PLD) is used to deposit layers of La$_{0.7}$Sr$_{0.3}$MnO$_{3}$ (LSMO) with a thickness of $30\,\rm{nm}$ on orthorhombic NdGaO$_{3}$(110) substrates (NGO). Subsequently without breaking the vacuum a layer of SrRuO$_{3}$ (SRO) with a thickness of $t_{\rm{NM}}$ is grown. Both materials are deposited using an oxygen pressure of $0.2\,\rm{mbar}$ at a substrate temperature of $650\,^{\circ}\rm{C}$. The laser fluency is $2.5\,\rm{J/cm}^2$ at a repetition rate of $5\,\rm{Hz}$. Using this recipe a total number of 7 samples is fabricated with different respective thickness $t_{\rm{NM}}$.
In addition two reference samples are fabricated, one single LSMO layer and one sample where the SRO is replaced by $8\,{\rm nm}$ of Pt which is also deposited without breaking the vacuum by magnetron sputtering. The samples are cleaved to approximately $2\,{\rm mm}$ times $5\,{\rm mm}$ rectangles where the substrate [001] direction is oriented along the short sides. On these sides copper leads are attached using silver glue to serve as voltage probes for the ISHE.
\subsection*{Measurement}
Samples are placed on top of a coplanar waveguide with a $600\,\mu\rm{m}$ wide inner conductor and $100\,\mu\rm{m}$ gap. The waveguide is electrically isolated by an approximately $100\,\rm{nm}$ thick polyimide layer. The waveguide is placed in a cryostat which is in constant magnetic field $\mu_{0}H_{\rm{ext}}$ of a rotatable electromagnet. A microwave current is transmitted through the waveguide whose RF magnetic field excites ferromagnetic resonance and thus spin precession in the LSMO layer. The cryostat is liquid nitrogen cooled allowing for temperature dependent measurement in the range from $300\,\rm{K}$ down to $80\,\rm{K}$. It is possible to modulate the RF amplitude and to use a lock-in amplifier to measure the ISHE or to apply a small AC magnetic field allowing for lock-in detection of the absorption and thus the ferromagnetic resonance.
\subsection*{Data analysis}
In analogue manner to \cite{Obstbaum2014}, the data from individual voltage measurements are fit to $V_{DC}=A_s\times L_s(H)+B_a\times L_a(H)+\rm{offset}$ with $L_s(H)={\Delta H}^2/\left(\left(H-H_r\right)^2+{\Delta H}^2\right)$ and $L_a(H)=\left(H-H_r\right)\Delta H/\left(\left(H-H_r\right)^2+{\Delta H}^2\right)$ to obtain the amplitudes for the symmetric and antisymmetric contribution of the voltage signal $A_s$ and $B_a$ respectively. For the case of $\varphi=0$ or $\varphi=180^\circ$, $A_s$ is equal to $V_{\rm ISHE}$ since any contribution from AMR becomes zero for this configuration. FMR measurements have been carried out subsequently to the voltage measurements from which the susceptibilities at resonance can be extracted.

\section*{Acknowledgements}
We thank Georg Woltersdorf for valuable discussions. This work was supported by the European Commission in the project IFOX under grant
agreement NMP3-LA-2010-246102 and by the Deutsche Forschungsgemeinschaft in the SFB762. .


\clearpage

\begin{figure}[ht]
\centering
\includegraphics[width=0.50\textwidth]{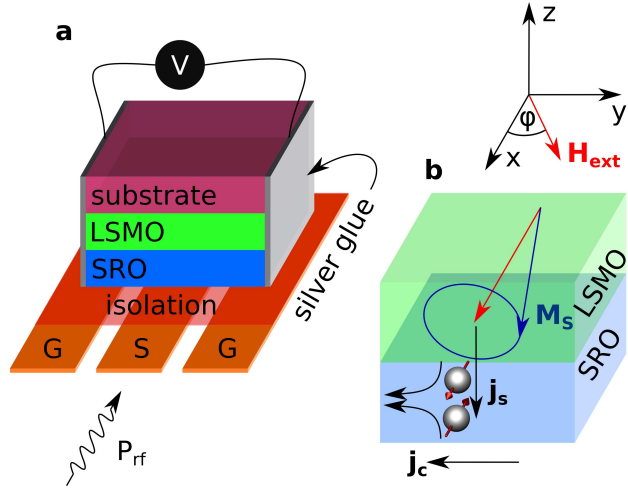}
\caption{\textbf{Measurement geometry. a }, Sample placed on waveguide separated by an electrically isolating coating. The generated voltage can be measured between the two contacts on the sample sides. \textbf{b}, Depiction of the ISHE voltage generation. A spin current $j_s$ generated by spin pumping from the LSMO into the SRO is converted to a charge current $j_c$.
\label{pic2}}
\end{figure}

\clearpage


\clearpage

\end{document}